\documentclass[useAMS,usecolumn,usenatbib,usegraphicx]{mn2e}
\usepackage{color}
\usepackage[varg]{txfonts}

\title[Detection of a turbulent gas associated with a starless core]{Detection of a turbulent gas component associated with a starless core with subthermal turbulence in the Orion A cloud}
\author[S. Ohashi et al.]{
\parbox[t]{\textwidth}{\vspace{-1cm}
Satoshi Ohashi,$^{\! 1,}$\thanks{E-mail: satoshi.ohashi@nao.ac.jp} Ken'ichi Tatematsu,$^{\! 2,3}$ Patricio Sanhueza,$^{\! 2}$ Quang Nguy$\tilde{\hat{\rm e}}$n Lu{\hskip-0.65mm\small'{}\hskip-0.5mm}o{\hskip-0.65mm\small'{}\hskip-0.5mm}ng,$^{\! 2}$ Tomoya Hirota,$^{\! 2,3}$ Minho Choi,$^{\! 4}$  and Norikazu Mizuno$^{1,2}$}\\\\
$^{1}$Department of Astronomy, The University of Tokyo, Bunkyo-ku, Tokyo 113-0033, Japan\\
$^{2}$National Astronomical Observatory of Japan, Osawa 2-21-1, Mitaka, Tokyo 181-8588, Japan\\
$^{3}$Department of Astronomical Science, SOKENDAI (The Graduate University for Advanced Studies), 2-21-1 Osawa, Mitaka, Tokyo 181-8588, Japan\\
$^{4}$Korea Astronomy and Space Science Institute,  Daedeokdaero 776, Yuseong, Daejeon 305-348, South Korea\\
}
\begin{document}

\date{Accepted ... ; Received ... ; in original form ...}

\pagerange{\pageref{firstpage}--\pageref{lastpage}} \pubyear{2012}

\maketitle

\label{firstpage}

\begin{abstract}
We report the detection of a wing component in NH$_3$ emission toward the starless core TUKH122 with subthermal turbulence in the Orion A cloud. This NH$_3$ core is suggested to be on the verge of star formation because the turbulence inside the NH$_3$ core is almost completely dissipated, and also because it is surrounded by CCS, which resembles the prestellar core L1544 in Taurus showing infall motions. Observations were carried out with the Nobeyama 45 m telescope at 0.05 km s$^{-1}$ velocity resolution. 
We find that the NH$_3$ line profile consists of two components. The quiescent main component has a small linewidth of 0.3 km s$^{-1}$ dominated by thermal motions, and the red-shifted wing component has a large linewidth of 1.36 km s$^{-1}$ representing turbulent motions. These components show kinetic temperatures of 11 K and $<$ 30 K, respectively.  Furthermore, there is a clear velocity offset between the NH$_3$ quiescent gas ($V_{\rm LSR}=3.7$ km s$^{-1}$) and the turbulent gas ($V_{\rm LSR}=4.4$ km s$^{-1}$). The centroid velocity of the turbulent gas corresponds to that of the surrounding gas traced by the $^{13}$CO ($J=1-0$) and CS ($J=2-1$) lines. LVG model calculations for CS and CO show that the turbulent gas has a temperature of $8-13$ K and an H$_2$ density of $\sim$ $10^4$ cm$^{-3}$, suggesting that the temperature of the turbulent component is also $\sim$ 10 K. The detections of both NH$_3$ quiescent and wing components may indicate a sharp transition from the turbulent parent cloud to the quiescent dense core. 
\end{abstract}

\begin{keywords}
ISM: individual (Orion Nebula, Orion Molecular Cloud)
---ISM: molecules
---ISM: structure---stars: formation
\end{keywords}

\section{Introduction}
``Giant molecular clouds (GMCs)'' are well known to be major sites of star formation in our Galaxy, and often show star cluster formation including massive stars \citep[e.g.,][]{shu87}. Molecular line observations of GMCs reveal that the observed linewidths are much larger than thermal linewidths and GMCs are known to contain supersonic turbulent motions \citep[e.g.,][]{lar81,sol87}.  ``Molecular clouds cores'' or ``dense cores'' having sizes of the order of 0.1 pc have been identified in GMCs and are thought to be birth places of stars \citep[e.g.,][]{mye83,bei86,lad91}. 
The turbulent motions still remain even in these dense cores and support the dense cores against gravitational collapse.
It has been suggested that the dissipation of turbulence can initiate the star formation process in turbulent cores \citep{nak98,mye98}.   However, it is poorly understood how turbulence is dissipated within cores and whether cores dissipate turbulence partially or completely before star formation. To investigate star formation within turbulent cores, it is essential to observationally characterize them in the turbulent environments where most stars are born.

\begin{figure*}
  \begin{center}
  \includegraphics[width=18cm]{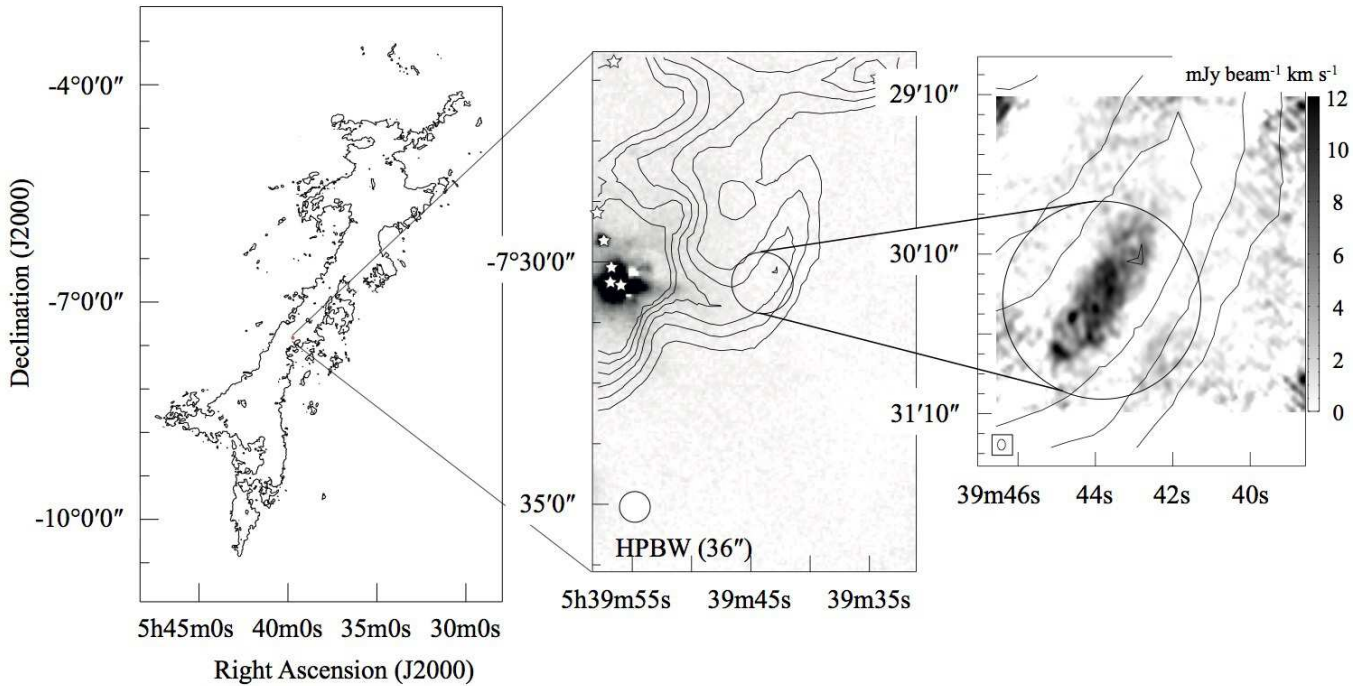}
  \end{center}
  \caption{ (Left):  The Orion A large map taken by $Herschel$ $SPIRE$ 500 $\mu$m dust emission is shown in a contour. The contour level is 13 mJy.  (Middle): The $Herschel$ $SPIRE$ 500 $\mu$m dust emission map in contour is imposed on the grey scale map of the $Herschel$ $PACS$ 70 $\mu$m emission. The lowest contour level is 92 mJy (10\% of the peak flux in the map) and the contour interval is 18.4 mJy (2\% of the peak flux in the map). The center circle represents the beam size of the Nobeyama 45 m telescope at 23 GHz.  The bottom-left circle is the $Herschel$ beam size (36$\arcsec$) at 500 $\mu$m. (Right): The NH$_3$ velocity-integrated intensity map in grey scale of the main NH$_3$ $(J,K)=(1,1)$ component toward TUKH122 obtained with VLA. The velocity integration range is from
$V_{\rm LSR}$ $=$ 3.2 to 4.4 km s$^{-1}$. The contours represent 500 $\mu$m dust continuum emission (same contour levels used in the left panel). The bottom-left circle is the VLA synthesized beam (4$\farcs1\times3\farcs1$). }
  \label{map}
\end{figure*}

We focus our study on dense cores located in the Orion A cloud, one of the nearest and best studied GMCs. The distance to the Orion A cloud is estimated to be 418 pc \citep{kim08}. At this distance, 1 arcmin corresponds to 0.12 pc.
The whole structure of the Orion A cloud is filamentary and contains ``$\int$-shaped filament'' in the northern region \citep{bal87,joh99}.  The $Spitzer$ $Space$ $Telescope$ carried out a large survey of the Orion A cloud with $IRAC$ and $MIPS$ instruments and provided a young stellar object (YSO) catalog \citep{meg12}.
Recently, the $Herschel$ $Space$ $Observatory$ carried out an extensive imaging survey toward the Orion A cloud with $SPIRE$ at $250-500$ $\mu$m and $PACS$ at $70-160$ $\mu$m. \citet{stu15} analyzed column density probability distribution function (N-PDF) and suggested that the N-PDF power-law index is associated with the evolutionary state of the gas.  These high sensitivity and large imaging surveys provide us with information on the physical condition of prestellar cores \citep[e.g.,][]{pol13}. To investigate the dynamical conditions of prestellar cores, molecular line observations are crucial. Wide field molecular line observations have been carried out toward the Orion A cloud using CO, $^{13}$CO, C$^{18}$O ($J=1-0$),  CS ($J=1-0$), NH$_3$ $(J,K)$ $=$ $(1,1)$ and $(2,2)$,  N$_2$H$^+$ ($J=1-0$), and others \citep{mad86,ces94,ike07,buc12,shi15,tat93,tat08}. 

 \citet{tat93} identified 125 molecular cloud cores in CS ($J=1-0$) emission and cataloged their positions (here we use the prefix of TUKH for their core numbers).
 Subsequently, \citet{tat10} made one-pointing observations toward more than 60 cores in the Orion A cloud in CCS, HC$_3$N, N$_2$H$^+$ and other lines. They detected CCS emission for the first time in the Orion A cloud. The carbon-chain molecules of CCS and HC$_3$N are thought to be tracers of chemically young molecular gas, while N-bearing molecules of N$_2$H$^+$ and NH$_3$ are thought to be tracers of evolved gas in nearby cold dark clouds \citep[e.g.,][]{suz92,ben98,hir02}.
\citet{tat14a}  mapped N$_2$H$^+$ ($J=1-0$) and CCS ($J_N=7_6-6_5$) toward six cores in the Orion A cloud with the Nobeyama 45 m telescope and confirmed that the N$_2$H$^+$/CCS column density ratio can be used as a chemical evolution tracer even in the Orion A cloud. Similarly,  \citet{oha14} found that the NH$_3$/CCS  column density ratio is anti-correlated with the CCS linewidth, and suggested that chemical evolution and turbulence dissipation can be indicators of the dynamical evolution of cores.
Among these cores, the prestellar core TUKH122 seems to be the closest to form stars because the N$_2$H$^+$/CCS column density ratio is the largest value among the starless cores.  \citet{tat14b} observed the NH$_3$ ($J,K$) $=$ $(1,1)$ and CCS ($J_N=4_3-3_2$) emission lines toward the TUKH122 core with the Very Large Array (VLA) and revealed narrow NH$_3$ line profiles ($\Delta v$ $\sim$ 0.2 km s$^{-1}$), in contrast to the broader previous observed CS ($J=1-0$) line profiles ($\Delta v$ $\sim$ 0.8 km s$^{-1}$). They also found that an NH$_3$ oval structure (core) is surrounded by CCS emission. This configuration is quite similar to that of  the starless cores L1544 and L1498 in Taurus \citep{aik01,lai00}.  The prestellar core L1544 displays infall motions \citep{taf98}, which means that NH$_3$ cores surrounded by CCS emission might indicate that these cores are chemically evolved and on the verge of the star formation.

Here, we report new single-pointing observations of the NH$_3$ ($J,K$) $=$ ($1,1$) and ($2,2$) emission lines toward the TUKH122 core at 0.05 km s$^{-1}$ velocity resolution using the Nobeyama 45 m telescope in order to investigate the kinematics and the physical conditions of the core. Measurements of the NH$_3$ inversion lines are ideal for temperature of dense gas \citep{ho83}.

\section{Observations}


Single-pointing observation was carried out with the Nobeyama 45 m telescope on 2015 March 4. 
At 23 GHz, the half-power beam width (HPBW) and main beam efficiency ($\eta$) of the telescope were 73$\arcsec$ and 0.825, respectively.
 The observations were perfomred in the position switch mode. For the receiver front end, we employed the 20 GHz HEMT receiver (H22). For the back end, we used the SAM45 digital spectrometer with a spectral resolution of 3.81 kHz (0.05 km s$^{-1}$ at 23 GHz)  and a bandwidth of 16 MHz. 
The NH$_3$ ($J,K)=(1,1$) and ($2,2$) lines were simultaneously observed.  We used rest frequencies of 23.69449 and 23.72263 GHz for NH$_3$ ($J,K)=(1,1$) and ($2,2$), respectively \citep{kuk67}.
The system noise temperatures ranged from 90 to 96 K. The standard
chopper wheel method was used, and the intensity is reported in terms of the main-beam temperature $T_{\rm mb}$, which is obtained by dividing the antenna temperature $T_{\rm A}^\star$ by the main beam efficiency.  The on-source integration time was $\sim$ 52 min, resulting in an rms noise level of 37 mK in $T_{\rm mb}$ for the NH$_3$ $(J,K)=(1,1)$ observations in the dual polarization mode, and 51 mK in $T_{\rm mb}$ for the NH$_3$ $(J,K)=(2,2)$ observations in the single polarization mode.  
The telescope pointing was checked at the beginning of the observation by observing the SiO maser source Orion KL and was better than 10$\arcsec$. 




\section{Results and Discussion}
 Figure \ref{map} shows the location of TUKH122 on top of 70 $\mu$m, thermal dust continuum map, and NH$_3$ emission map.  The bright 70 $\mu$m source corresponds to TUKH123, which contains six protostars identified by $Spitzer$ \citep{meg12}. On the other hand, no 70 $\mu$m emission was detected toward the TUKH122 core, which is consistent with the fact that the TUKH122 core has no protostars. We can see a filamentary structure elongated from northwest to southeast toward the TUKH122 core (Figure \ref{map}). The width of the filament is  about 0.1 pc and its length is 0.5 pc.  On the NH$_3$ map, an oval structure with embedded condensations is seen toward TUKH 122.

\subsection{Hyperfine fitting and derivation of physical parameters}

\begin{figure*}
  \begin{center}
  \includegraphics[width=14cm]{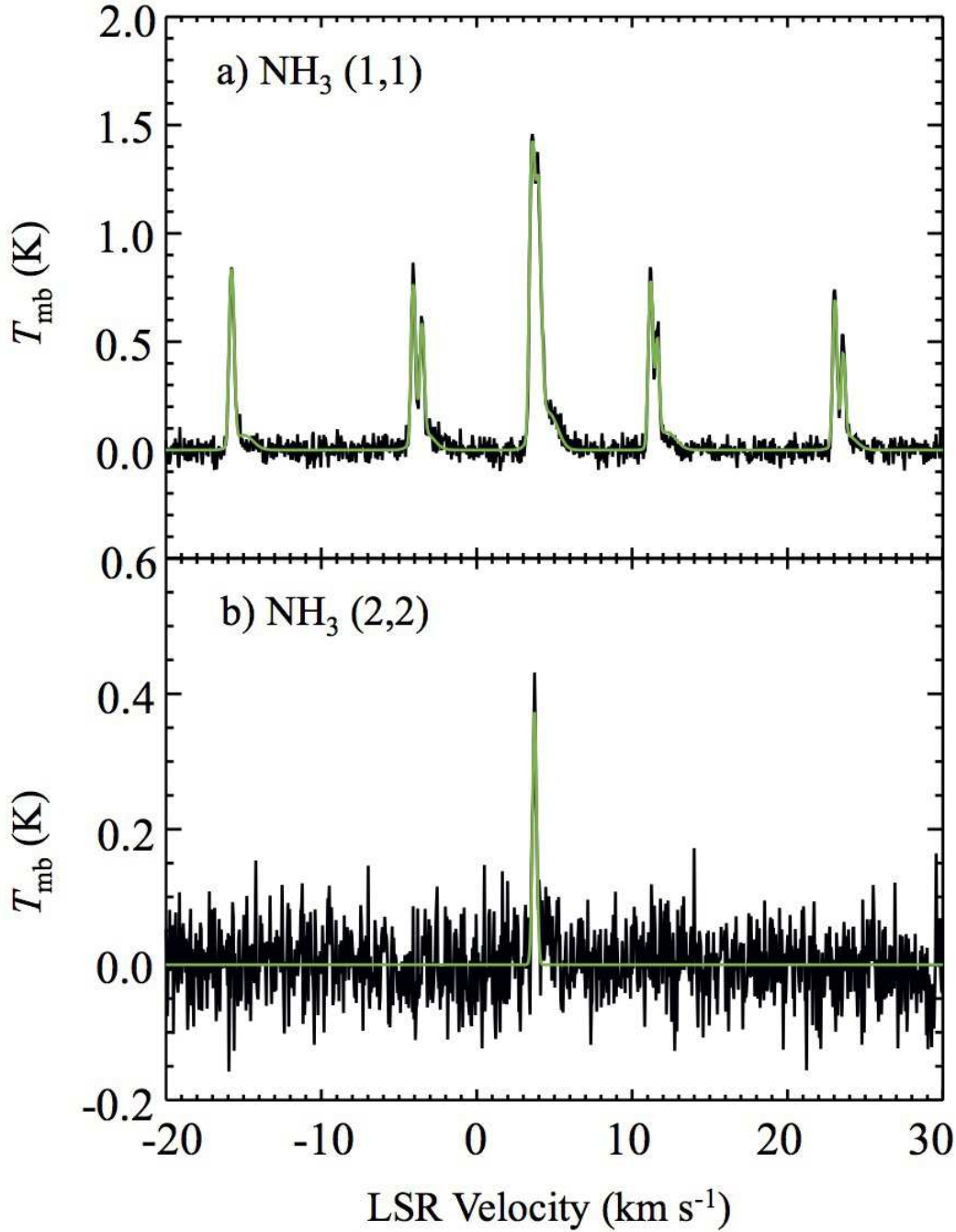}
  \end{center}
  \caption{(a) The NH$_3$ $(J,K)=(1,1)$ spectrum and hyperfine-fitting result. The green line represents the best fit model of the hyperfine fitting.  (b) The NH$_3$ $(J,K)=(2,2)$ spectrum and Gaussian-fitting result. The green line represents the best fit model of the Gaussian-fitting.}
  \label{nh3spectra}
\end{figure*}

Figure \ref{nh3spectra} shows the line profile and hyperfine fitting of the NH$_3$ ($J,K$) $=$ ($1,1$) and ($2,2$) lines obtained with the Nobeyama 45 m telescope. We identified a total of nine hyperfine components with $T_{\rm mb}$ $\sim$ $0.8-1.5$ K in the NH$_3$ ($J,K$) $=$ ($1,1$) line. 
 The hyperfine fitting was performed to the NH$_3$ ($J,K=1,1$) spectra by using equations:
 \begin{equation}
T_{\rm A}^\star(v) = \eta[J_{\nu}(T_{\rm ex}) - J_{\nu}(T_{\rm bg})][1 - \exp(-\tau (v))],
\label{1}
\end{equation}
where
\begin{equation}
\tau(v) = \tau_{\rm tot}\sum_i \frac{s_i}{4.284} \exp\left\{-4\ln2 \left(\frac{v-v_i-v_{\rm LSR}}{\Delta v}\right)^2\right\}.
\label{2}
\end{equation}
The optical depth $\tau_{\rm tot}$ is the sum of the centroid optical depths of all the hyperfine components.  The background temperature $T_{\rm bg}$ is 2.7 K.  $J_{\nu}$ is the Planck function in temperature units.
 The free parameters are the optical depth ($\tau_{\rm tot}$), LSR velocity ($V_{\rm LSR}$), linewidth ($\Delta v$), and excitation temperature ($T_{\rm ex}$). We assume a uniform excitation temperature in all NH$_3$ hyperfine components. The intrinsic line strengths ($s_i$) and velocity offsets ($v_i$) of the hyperfine components are adopted from \citet{lev10}.   We fit the hyperfine structure with a least-squares Gaussian hyperfine structure fitting routine written in IDL. Fitting is optimized by the use of the Levenberg-Marquardt algorithm (for non-linear fitting).
Iterations are performed until the solutions converge.

Because we identified another component (wing component) in the NH$_3$ ($J,K$) $=$ ($1,1$) spectrum, we fit a two-velocity-component hyperfine model. The fitting results of the wing emission will be discussed in the following subsection.
The fitting results of the quiescent component are: $\tau_{\rm tot}=8.4\pm0.1$, $V_{\rm LSR}=3.74\pm$0.01 km s$^{-1}$, $\Delta v=0.30\pm0.001$ km s$^{-1}$, and $T_{\rm ex}=4.14\pm0.05$ K.
The NH$_3$ hyperfine components are moderately optically thick. It is found that the core is thermally dominated because the observed linewidth is close to the thermal linewidth $\Delta v$(th) $=$ $(8\ln2\hspace{1mm}kT_{\rm k}/m)^{1/2}$, which is 0.16 km s$^{-1}$ for NH$_3$ at $T_{\rm k}=10$ K. Here, $k$ is the Boltzmann constant, and $m$ is the mass of the molecule.

We performed a single Gaussian fitting to the NH$_3$ ($J,K$) $=$ ($2,2$) spectrum because there is only one detected velocity component and hyperfine lines are not detected.  The fitting results are: peak intensity of $T_{\rm mb}=0.37\pm0.03$ K, centroid velocity of $V_{\rm LSR}=3.728\pm0.014$ km s$^{-1}$, and linewidth of $\Delta v=0.31\pm0.03$ km s$^{-1}$.
The optical depth of $\tau(2,2)$ can be derived from equation (1) assuming that NH$_3$ ($J,K$) $=$ ($1,1$) and ($2,2$) transition have the same excitation temperature and a filling factor of unity for both transitions.
The rotational temperature of NH$_3$ was derived from the ($J,K$) $=$ ($1,1$) and ($2,2$) transitions by following \citet{bac87} and equation (\ref{3})
\begin{equation}
T_{\rm rot}=\frac{41.5}{\ln\Big(2.35\frac{\tau_m(1,1)}{\tau_m(2,2)}\frac{\Delta v_{1,1}}{\Delta v_{2,2}}\Big)},
\label{3}
\end{equation}
where $\tau_m(1,1)$ is the optical depth of the main hyperfine component group ($\tau_m(1,1)=0.5\times\tau_{\rm tot}$).
Following \citet{man92} and \citet{man15}, the column density was derived as

\begin{displaymath}
N(J,K)=3.96\times10^{12}\frac{J(J+1)}{K^2}\hspace{50mm}
\end{displaymath}
\begin{equation}
\times\:\Bigg\{\frac{1+\exp(-h\nu/kT_{\rm ex})}{1-\exp(-h\nu/kT_{\rm ex})}\Bigg\}\tau(J,K)\Delta v \hspace{2mm}{\rm cm^{-2}}.
\label{4}
\end{equation}
The total column density was obtained by using
\begin{equation}
N({\rm NH_3})=0.0138\times N(1,1)\exp\Bigg(\frac{23.1}{T_{\rm rot}}\Bigg) T_{\rm rot}^{3/2}.
\label{5}
\end{equation}
From these equations, we derived a rotation temperature of $T_{\rm rot}=10.6\pm1.5$ K and $N({\rm NH_3})=(5.8\pm0.5)\times10^{14}$ cm$^{-2}$.
In order to estimate the kinetic temperature $T_{\rm kin}$, we used the relations derived by \citet{taf04}
\begin{equation}
T_{\rm kin}=\frac{T_{\rm rot}}{1-\frac{T_{\rm rot}}{42}\ln\Big[1+1.1\exp(-16/T_{\rm rot})\Big]}
\label{6}
\end{equation}

The kinetic temperature is derived to be $T_{\rm kin}=11\pm2$ K. The value of $T_{\rm kin}=11$ K is consistent with the fact that there is no
heating source in this core.
We also investigate how the filling factor affects equation (1). If the filling factor is 0.5, the kinetic temperature and the column density are derived to be $T_{\rm kin}$ $=$ 11 K and $N({\rm NH_3})=7.8\times10^{14}$ cm$^{-2}$, respectively. The kinetic temperature is not significantly affected by the filling factor because it is derived from the intensity ratio of NH$_3$ $(J,K$) $=$ ($1,1$) and $(2,2)$ emission.  Again, we assume that both transitions have the same filling factor.

\subsection{Wing emission}
\begin{figure}
  \begin{center}
  \includegraphics[width=8cm]{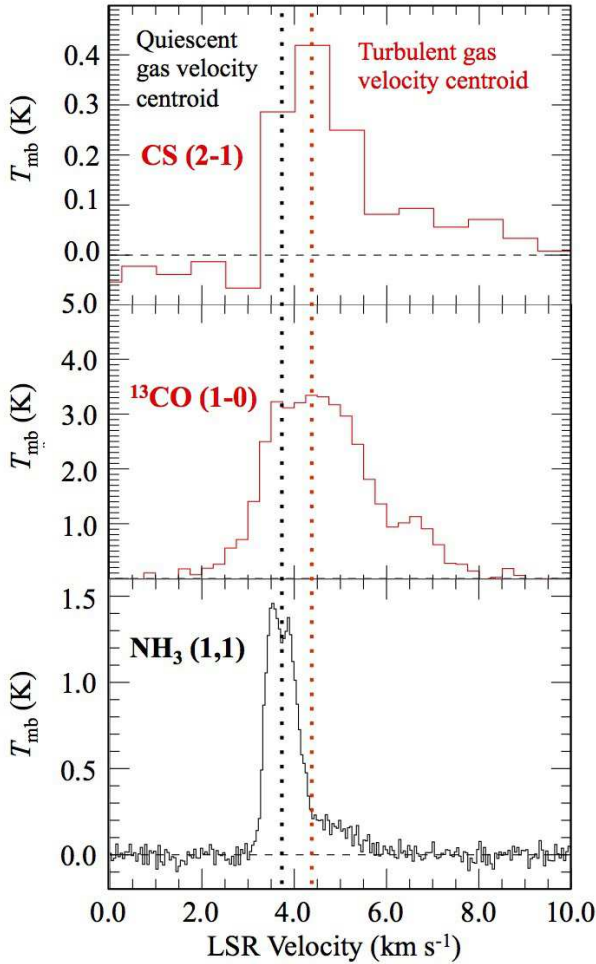}
  \end{center}
  \caption{ CS ($J=2-1$), $^{13}$CO($J=1-0$), and NH$_3$ $(J,K=1,1)$ line profiles. The black dotted line indicates the centroid velocity of the quiescent NH$_3$ gas at 3.7 km s$^{-1}$ and the red dotted line indicates the turbulent NH$_3$ gas at 4.4 km s$^{-1}$.   }
  \label{profile}
\end{figure}

In addition to the narrow linewidth component of $\Delta v=0.3$ km s$^{-1}$, we identified a red-shifted wing component in the NH$_3$ $(J,K)$ $=$ $(1,1)$ spectrum for the first time for this source (see Figures \ref{nh3spectra} a and \ref{profile}). 

To analyze the wing emission, we performed the hyperfine fitting including the quiescent main component and the wing component, simultaneously.
The fitting results of the wing component are: $\tau_{\rm tot}=3.23\pm0.03$, $V_{\rm LSR}=4.40\pm$0.02 km s$^{-1}$, $\Delta v=1.36\pm0.01$ km s$^{-1}$, and $T_{\rm ex}=2.9\pm0.1$ K, suggesting that the wing emission is turbulent.  Because we were unable to detect a counterpart in the NH$_3$ ($J,K$) $=$ $(2,2$) emission, we estimated the upper limit to the kinetic temperature to be 30 K assuming the 3 $\sigma$ upper limit of the NH$_3$ ($J,K$) $=$ $(2,2$) emission. 
The centroid velocity of NH$_3$ quiescent gas is 3.7 km s$^{-1}$, while that of the wing/high velocity gas is 4.4 km s$^{-1}$.  The velocity offset is about 0.7 km s$^{-1}$,  more than twice the linewidth of the quiescent component. 
In Figure \ref{profile}, we show the  CS ($J=2-1$), $^{13}$CO ($J=1-0$), and NH$_3$ ($J,K=1,1$) line profiles toward TUKH122.  The $^{13}$CO ($J=1-0$) and CS ($J=2-1$) lines are convolved with a Gaussian kernel to match the beam size of the NH$_3$ observations (73$\arcsec$). These lines were observed by \citet{tat93,tat98,tat14a}. We find that the $^{13}$CO ($J=1-0$) and CS ($J=2-1$) emission lines are dominated by turbulent motions and their velocity centroids are consistent with that of the NH$_3$ turbulent gas rather than that of the NH$_3$ quiescent gas within the uncertainties (see also table 1). 
A rest frequency of 110.201353 GHz was used for $^{13}$CO ($J=1-0$) \citep{uli76}.

\begin{table}
 \centering
 \caption{ Fitting results of various molecular lines toward TUKH122}
 \resizebox{\columnwidth}{!}{%
   \begin{tabular}{ccccc}
     \hline
      molecular lines & $\tau$  & $V_{\rm LSR}$     &  $\Delta v$ &$T_{\rm ex}$  \\
            &   &  (km s$^{-1}$) & (km s$^{-1}$) & (K)  \\
     \hline
     NH$_3$ ($J,K=1,1$)   main     & 8.4$\pm$0.1   & 3.74$\pm$0.01 &  0.30 & 4.14$\pm$0.05\\
     NH$_3$ ($J,K=1,1$)   wing     & 3.23$\pm$0.03   & 4.40$\pm$0.02 &  1.36 & 2.9$\pm$0.1\\
     CS ($J=2-1$)     &    & 4.37$\pm$0.33 &  1.7 & \\
     $^{13}$CO ($J=1-0$)     &   & 4.51$\pm$0.09 &  2.6 & \\
     
           \hline     \label{tab:fitting}
   \end{tabular}   }
\end{table}

In addition to the upper limit to the kinetic temperature of the turbulent gas explained above, we tried to obtain the temperature by another method. We make LVG calculations \citep[e.g.,][]{sco74,gol74} for the multitransition CS $(J=1-0)$, CS($J= 2-1)$ and C$^{34}$S $(J=2-1)$ data to derive the kinetic temperature of the turbulent gas.  LVG analysis was made by using the RADEX software \citep{van07}.  The collision rates for CS are taken from \citet{liq06}. We assume that the CS lines are emitted from similar volumes and they are from a uniform density sphere. 

\begin{figure}
  \begin{center}
  \includegraphics[width=8cm]{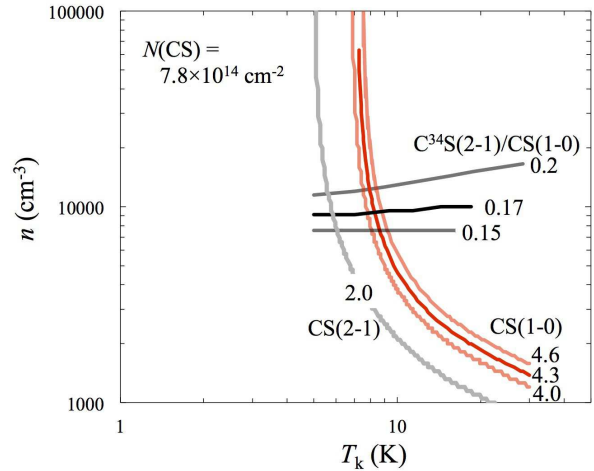}
  \end{center}
  \caption{ LVG results for $N$(CS)=7.8$\times 10^{14}$ cm$^{-2}$ and $\Delta v=$0.56 km s$^{-1}$ are shown in the density-temperature diagram. Black, red, and grey lines show the C$^{34}$S ($J=2-1$)/CS($J=1-0$) intensity ratio, CS ($J=1-0$) intensity, and CS($J=2-1$) intensity, respectively.}
  \label{lvg}
\end{figure}

 In LVG calculations, we investigate the parameter range of $n(\rm H_2)=10^3-10^5$ cm$^{-3}$ and $T_{\rm kin}=5-30$ K. We use a column density of $N$(CS) $\sim$ $7.8\times10^{14}$ cm$^{-2}$, obtained from CS $(J=2-1)$ by assuming LTE condition, and a linewidth of 0.56 km s$^{-1}$. 
The formulation can be found, for example, in \citet{lee13}.  We assume the abundance ratio of $\rm ^{32}S/^{34}S$ to be 15 \citep{bla87}. We use a CS ($J=1-0$) intensity of 4.3, a C$^{34}$S ($J=2-1$)/CS($J=1-0$) intensity ratio of 0.17, and a CS ($J=2-1$) intensity of 0.7  toward the peak position of TUKH122.
 Figure \ref{lvg} shows the results and the derived values are listed in Table 2.  Temperature is significantly low ($\sim$ 10 K) even taking into account the uncertainties of the intensities. The observed CS $(J=2-1)$ intensity is generally weaker than the value determined by the model. This has been already pointed out by \citet{tat98}. They suggested that foreground absorption due to less dense gas is significant for $J=2-1$. \citet{nis15} also performed LVG analysis using CO $(J=1-0)$, CO $(J=2-1)$, $^{13}$CO $(J=1-0$), and $^{13}$CO $(J=2-1)$ lines and derived a kinematic temperature of $\sim$ 13 K toward TUKH122. Therefore, we conclude that the turbulent gas is also cold ($\sim$ 10 K).
 
 \begin{table}
 \centering
 \caption{ LVG models for CS toward TUKH122}
 \resizebox{\columnwidth}{!}{%
   \begin{tabular}{ccccc}
     \hline
      model & $n$  & $T_{\rm k}$     &  $N$(CS) &Min  $\chi^2$  \\
            & (cm$^{-3}$)  &  (K) & (cm$^{-2}$) &  \\
     \hline
     fitting range       & $10^3-10^5$   & $5-30$ &  7.8$\times10^{14}$ & \\
     results       & 8.7$\times10^3$   & 8 &  7.8$\times10^{14}$ & 0.0002\\
           \hline
     \label{tab:lvg}
   \end{tabular}   }
\end{table} 
 
It is worth noting that our previous results of NH$_3$ observations with the VLA \citep{tat14b} have shown no wing emission toward the core. The VLA observations were carried out using NH$_3$ ($J,K$) $=$ ($1,1$) emission with 0.2 km s$^{-1}$ velocity resolution (Figure 7 in Tatematsu et al. 2014b). 
From the observed intensity of the wing emission in the present study ($T_{\rm mb}$ $=$ 0.22 K), the flux density is estimated to be 1.31 mJy beam$^{-1}$ at the VLA configuration of 4$\farcs$2 $\times$ 3$\farcs$1 resolution. The wing component were not detected with the VLA because the rms noise level of the VLA observations was as high as $\sim$ 2.6 mJy beam$^{-1}$. Therefore, the non-detection of the wing component with the VLA will be due to lower sensitivity.  Another possibility is that the turbulent gas is widely extended and resolved out in the interferometric observations.

\subsection{The linewidth-size relations}
The linewidth-size relation has been investigated in many studies \citep[e.g.,][]{lar81,cas95,goo98}. It is suggested that the coefficient or intercept (in the log-log form) of the linewidth-size relation in GMC turbulent cores is larger than in low-mass star-forming regions, which means GMC cores have a higher level of turbulence \citep{tat93,cas95,bal11,hey09}.  \citet{lu14} investigated the linewidth-size relation in high-mass star-forming regions in NH$_3$ emission using the VLA. Their results suggest that high-mass star-forming cores still have high level of turbulence within 0.1 pc.

We estimate the nonthermal linewidth $\Delta v_{\rm NT}$, which is defined as $\Delta v_{\rm NT}^2= \Delta v_{\rm obs}^2-\Delta v_{\rm T}^2$, where $\Delta v_{\rm T}=(8\ln2\hspace{1mm}kT_{\rm k}/m_{\rm obs})^{1/2}$ and $m_{\rm obs}$ is the mass of the observed molecules \citep{ful92}. We assume $T_k$ $=$ 11 K, which was obtained from the NH$_3$ inversion transition observations, for all molecular lines. 
The thermal linewidth for NH$_3$ and mean molecular weight (2.33 u) are 0.17 km s$^{-1}$ and 0.47 km s$^{-1}$, respectively.

\begin{figure}
  \begin{center}
  \includegraphics[width=8cm]{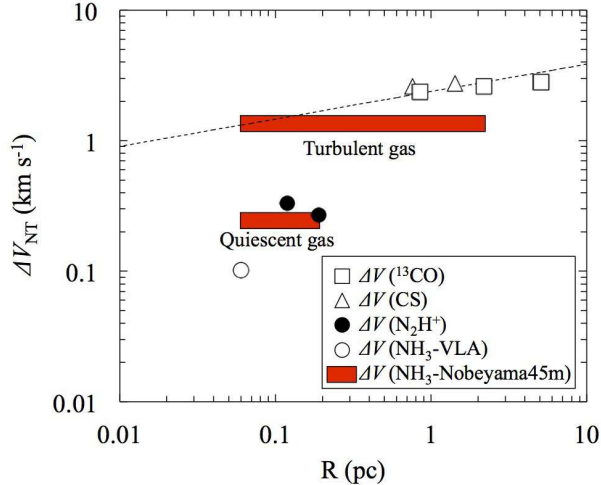}
  \end{center}
  \caption{ The nonthermal linewidth traced by different molecules against the core radius of the TUKH122 core. Values were derived by using the $^{13}$CO ($J=1-0$), CS ($J=2-1$), N$_2$H$^+$ ($J=1-0$), and NH$_3$ ($J,K=1,1$) observations. The dashed line represents a power law index of 0.21 derived by \citet{cas95}. }
  \label{r-dv}
\end{figure}

Figure \ref{r-dv} shows the nonthermal linewidth against the core radius in TUKH122 observed in several molecular lines; $^{13}$CO ($J=1-0$), CS ($J=2-1$), NH$_3$ ($J,K$) $=$ $(1,1$), and N$_2$H$^+$ ($J=1-0$).  The (equivalent) core radius is calculated from the area inside the 25\%, 50\%, and 75\% levels with respect to the maximum intensities of these molecular lines.  That is the core radius can be represented
\begin{equation}
r(25, 50, 75 \%)=\sqrt{S(25, 50, 75 \%)/\pi}
\label{8}
\end{equation}
We derived the observed linewidth, $\Delta v_{\rm obs}$, from the average spectrum from positions within the contours. We omit the plot of $r$(75\%) for CS and N$_2$H$^+$ because $S$(75\%) for CS and N$_2$H$^+$ is similar  to the telescope beam area and the contour is not fully resolved.  Our NH$_3$ single dish observations cannot constrain the radius. Therefore, we use the horizontal bar to illustrate the probable radius range of the NH$_3$ quiescent core. The NH$_3$ quiescent core was observed with VLA, but it was found that diffuse emission was resolved out. It is possible that the NH$_3$ quiescent core is  as large as the N$_2$H$^+$ quiescent, if N$_2$H$^+$ and NH$_3$ coexist. We also use the horizontal bar for the NH$_3$ turbulent core radius. The NH$_3$ turbulent gas distribution can be as small as the NH$_3$ quiescent core observed with VLA, while it can be widespread as observed in CS and $^{13}$CO with the Nobeyama 45 m telescope. 
The dashed line represents a power law index of 0.21 derived by \citet{cas95} in Orion cores. We find that the nonthermal component in TUKH122 follows the trend in the Orion cores if the core radius is larger than $\sim$ 0.2 pc. We also find that turbulence is dissipated within $\sim$ 0.2 pc, and the nonthermal linewidth becomes smaller than the thermal linewidth (0.47 km s$^{-1}$) for the mean molecular weight. 
 Our new results with high velocity resolution confirm that the nonthermal motion is almost completely dissipated. This may be the first evidence of a coherent region in a GMC core.

The nonthermal linewidth $\Delta v_{\rm NT}$ of the NH$_3$ wing component is derived to be 1.36 km s$^{-1}$, which is consistent with the trend of  the linewidth-size relation derived for CS and $^{13}$CO turbulent gas. Furthermore, the centroid velocity of the NH$_3$ wing component of 4.7 km s$^{-1}$ corresponds to that of CS and $^{13}$CO lines. Therefore, the NH$_3$ wing component should also trace the turbulent surrounding gas.  We have observed dense turbulent gas and quiescent dense gas in a single tracer. Our detection of both NH$_3$ quiescent and wing components may indicate a sharp transition from the turbulent parent cloud to the quiescent high dense core.

Similar objects that have a sharp transition between turbulent surrounding gas and a coherent core have been reported by \citet{pin10,pin11}.
However, it is still not clear how these coherent cores are formed in such turbulent environments.
Some models suggest that shocks may dissipate the turbulence \citep{pon12}.
Recently, hydrodynamic turbulent cloud simulations show the filament formation made up of a network smaller and coherent sub-filaments \citep{smi15}.
They found that sub-filaments are formed at the stagnation points of the turbulent velocity fields where shocks dissipate the turbulence.
Our results of the velocity offsets between the quiescent dense core and its parent cloud may be consistent with such simulations and TUKH122 core might be formed at the stagnation point.



\section{Conclusion}

We observed NH$_3$ ($J,K$) $=$ $(1,1$) and ($2,2$) lines toward the quiescent starless core TUKH122 with the Nobeyama 45 m telescope. 
We found that the NH$_3$ ($J,K$) $=$ $(1,1$) line profile consists of not only a quiescent component ($\Delta v$ $\sim$ 0.30 km s$^{-1}$) but also a wing turbulent component ($\Delta v$ $\sim$ 1.36 km s$^{-1}$).
 Analyzing the NH$_3$ ($J,K$) $=$ $(1,1$) and ($2,2$) lines, we derived kinetic temperatures of 11 K and $<$ 30 K for the  quiescent and turbulent components, respectively. Taking into account the fact that the turbulence is rapidly dissipated within a $\sim$ 0.2 pc region, this core may have a coherent region. This is the first detection of such an object in the GMC.
There is a clear offset between the centroid velocity of NH$_3$ quiescent gas ($V_{\rm LSR}=3.7$ km s$^{-1}$) and that of the turbulent gas ($V_{\rm LSR}=4.4$ km s$^{-1}$). The centroid velocity of the turbulent gas, rather than the quiescent one, corresponds to that of the surrounding gas traced by the $^{13}$CO ($J=1-0$) and CS ($J=2-1$) lines.
Our results from the NH$_3$ high velocity resolution observations may indicate a sharp transition from the turbulent parent cloud to the quiescent high dense core.

\section*{Acknowledgments}

We appreciate the anonymous referee for the very helpful comments to improve this paper. 
We are grateful to the staff of Nobeyama Radio Observatory. Nobeyama Radio
Observatory is a branch of the National Astronomical Observatory of Japan, National
Institutes of Natural Sciences.
Data analysis were carried out on common use data analysis computer system at the Astronomy Data Center, ADC, of the National Astronomical Observatory of Japan.

\label{lastpage}

\end{document}